\documentclass{book}
\usepackage{maxent,xspace,graphicx}

\def\bm#1{\mbox{\boldmath $#1$}}
\def\bb{{\bm b}}

\def\fb{{\bm f}}
\def\gb{{\bm g}}
\def\hb{{\bm h}}

\def\mb{{\bm m}}

\def\ub{{\bm u}}
\def\vb{{\bm v}}

\def\xb{{\bm x}}
\def\yb{{\bm y}}

\def\Ab{{\bm A}}

\def\Db{{\bm D}}

\def\Ib{{\bm I}}

\def\Mb{{\bm M}}

\def\Qb{{\bm Q}}

\def\thetab{\bm{\theta}}

\def\lambdab{\bm{\lambda}}

\def\psib{\bm{\psi}}

\def\Sigmab{\bm{\Sigma}}

\def\wh#1{\widehat{#1}}
\def\Lra{\Longrightarrow}

\def\d#1{\,\mbox{d}#1}

\def\disp#1{{\displaystyle #1}}

\def\intd{\int\kern-.8em\int}
\def\intt{\int\kern-.8em\int\kern-.8em\int}
\def\intg{\int\kern-1.1em\int}
\def\expf#1{\exp\left[ {#1} \right]}

\def\dpdx#1#2{{{\partial {#1}\over \partial {#2}}}}

\def\argmins#1#2{\mbox{arg}\min_{#1}\left\{{#2}\right\}}
\def\argmaxs#1#2{\mbox{arg}\max_{#1}\left\{{#2}\right\}}

\def\argmax#1#2{\mathop{\mbox{arg}\max}_{#1}\left\{{#2}\right\}}

\def\espx#1#2{\mbox{E}_{#1}\left\{ #2 \right\}}

\def\uncatcodespecials{\def\do##1{\catcode`##1=12 }\dospecials}

\newcount\lineno
\def\setupverbatim{\tt \lineno=0
 \obeylines \uncatcodespecials \obeyspaces
 \everypar{\advance\lineno by1 \llap{\sevenrm\the\lineno\ \ }}}
{\obeyspaces\global\let =\ }

\def\ER{\mbox{I\kern-.25em R}}
\def\EC{\mbox{C\kern-.8em C}}
\def\EZ{\mbox{Z\kern-.55em Z}}
\def\EN{\mbox{N\kern-.8em N}}

\def\beqnarr#1&#2&#3\\#4&#5&#6\eeqnarr{
    \left\{
           \begin{array}{lcl}
            {\displaystyle #1} & #2 & {\displaystyle #3} \\ 
            {\displaystyle #4} & #5 & {\displaystyle #6} 
           \end{array}
    \right. }

\def\ie{{\em i.e.}}

\def\beq{\vspace{-6pt}\begin{equation}}
\def\eeq{\end{equation}}
\def\beqn{\vspace{-12pt}\begin{eqnarray}}
\def\eeqn{\end{eqnarray}}

\newtheorem{Definition}{Definition}
\newtheorem{Theorem}{Theorem}
\newtheorem{Lemme}{Lemma}

\title{SCALE INVARIANT MARKOV MODELS FOR BAYESIAN INVERSION OF LINEAR INVERSE PROBLEMS}

\author{St\'ephane Brette, J\'er\^ome Idier and Ali Mohammad-Djafari \\
Laboratoire des Signaux et Syst\`emes (CNRS-ESE-UPS) \\ 
\'Ecole Sup\'erieure d'\'Electricit\'e, \\ 
Plateau de Moulon, 91192 Gif-sur-Yvette Cedex, France}

\begin{document}
\maketitle
\thispagestyle{empty}

\begin{abstract}
In a Bayesian approach for solving linear inverse problems one needs to 
specify the prior laws for calculation of the posterior law.
A cost function can also be defined in order to have a common tool 
for various Bayesian estimators which depend on the data and the hyperparameters. 
The Gaussian case excepted, these estimators are not linear and so depend on  
the scale of the measurements. 
In this paper a weaker property than linearity is imposed on the Bayesian estimator, 
namely the scale invariance property (SIP).

First, we state some results on linear estimation and then we introduce  
and justify a scale invariance axiom. We show that arbitrary choice 
of scale measurement can be avoided if the estimator has  this SIP. 
Some examples of classical regularization 
procedures are shown to be scale invariant. 
Then we investigate general conditions on classes of 
Bayesian estimators which satisfy this SIP, as well as their 
consequences on the cost function and  prior laws. 
We also show that classical methods for hyperparameters estimation 
({\em i.e.}, Maximum Likelihood  and Generalized Maximum Likelihood) 
can be introduced for hyperparameters estimation, and we verify the  
SIP property for them.

Finally we discuss how to choose the prior laws to obtain scale 
invariant Bayesian estimators. For this, we consider two cases of prior laws~: 
{\em entropic prior laws} and  {\em first-order Markov models}.
In related preceding works \cite{Mohammad-Djafari90b,Mohammad-Djafari93c}, 
the SIP constraints have been studied for the case of entropic prior laws. 
In this paper extension to the case of first-order Markov 
models is provided. 

\smallskip\noindent{\bf KEY WORDS :} 
Bayesian estimation, Scale invariance, Markov modelling, Inverse Problems, 
Image reconstruction, Prior model selection
\end{abstract}

\section{Introduction}

Linear inverse problem is a common framework for many different objectives, such as  
reconstruction, restoration, or deconvolution of images arising in 
various applied areas \cite{Demoment89a}.
The problem  is to  estimate  an object $\xb$ which is indirectly observed through 
a linear operator $A$, and is therefore noisy. We choose explicitly this
linear model because its simplicity captures
many of interesting  features of more complex 
models without their computational complexity.
Such a degradation models allows the following description: 

\beq
 \yb= \Ab \xb + \bb ,
\eeq 
where $\bb$ includes both the modeling errors and unavoidable
noise of any physical observation system, and $\Ab$ represents 
the indirect observing system and depends on a particular application.
For example, $\Ab$ can be diagonal or block-diagonal in deblurring,
Toeplitz or bloc-Toeplitz in deconvolution, or have no special interesting 
form as in X-ray tomography.

In order to solve these problems, one may choose to minimize the
quadratic  residual error $\disp{\|\yb - \Ab \xb\|^2}$. 
That leads to the classical linear system 

\beq
 \Ab^t \Ab \wh{\xb} =\Ab^t \yb .
\eeq
When mathematically exact solutions exist, they are too sensitive to
unavoidable noise and so are not of practical interest. 
This fact is due to a very high condition number of $A$ \cite{Demoment89a}.
In order to have a solution of interest, we must mathematically qualify admissible solutions. 

The Bayesian framework is well suited for this kind of problem because it could
combine information from data $\yb$  and prior knowledge on the solution.
One needs then to specify the prior laws $p_x(\xb;\lambdab)$ and 
$p_b(\yb-\Ab \xb;\psib)$ for calculation of the posterior 
$p_{x|y}(\xb|\yb) \propto p_x(\xb) \, p_b(\yb-\Ab \xb)$ with the Bayes rules.
Most of the classical Bayesian estimators, {\it e.g.}, Maximum {\em a posteriori} 
(MAP), Posterior Mean (PM) and Marginal MAP (MMAP), can be studied using 
the common tool of defining a cost function 
$C(\xb^*,\xb)$ for each of them. It leads to the classical Bayesian
estimator

\beq
 \wh{\xb}(\yb,\thetab)  
 = \argmins{\xb}{\espx{\xb^*|\yb}{C(\xb^*,\xb)|\yb }} 
\eeq
depending both on data $\yb$ and hyperparameters $\thetab$.

Choosing a prior model is a difficult task. This prior model would
include our prior knowledge. Some criteria based on information theory
and maximum entropy principle, have been used for that. For example, 
when our prior knowledge are the moments of the image   to be restored,
application of maximum entropy principle leads {\sc Djafari \& Demoment} 
\cite{Mohammad-Djafari90a} to exact determination
of the prior, including its parameters.  Knowledge of the bounds (a gabarit) and
the choice of a reference measure leads
{\sc LeBesnerais} \cite{LeBesnerais91a,LeBesnerais91b} to the construction of a
model accounting for human shaped prior in the context of astronomic
deconvolution. 

We consider the case when there is no important and quantitative 
prior information such as the knowledge of moment or bounds of the
solution. Then we propose to reduce the arbitrariness of the choice 
of prior model by application of constraint to the resulting Bayesian estimator. 
The major constraint for the estimator
is to be scale invariant, that is, whichever the scale or physical unit we choose, 
estimation results must be identical.
This desirable property will reduce the possible choice for prior
models and make it independent of the unavoidable scale choice. 
In this sense, related works of {\sc Jaynes} \cite{Jaynes68} or 
{\sc Box \&  Tiao}  \cite{Box72} on non-informative prior are 
close to our statement, although 
in these works the ignorance is not limited to the measurement scale. 
In our work, qualitative information only is supposed to be known 
(positivity excepted), so we think of choosing a parametric family of probability laws
as a usual and natural way in accounting for the prior. 
The parameters estimation in the chosen family of laws will be done according to the data, 
with a Maximun Likelihood (ML) or the Generalized Maximum Likelihood (GML) approach. 
These approaches are shown in this paper to be scale invariant. 

One can criticize choosing the prior law from a desired property of the final estimator 
rather than from the available prior knowledge. 
We do not maintain having exactly chosen a model but just restricting 
the available choice. 
Then  Gaussian or convex prior popularity
is  due likely to the tractability of the associated estimator rather than
Gaussianity or convexity of the modeling process. 
Lastly, good as the model is, its use depends on the
tradeoff between the good behavior of the final estimator and the quality of estimation.

The paper is organized as follows. First, we state some known results on
Gaussian estimators as well as introduce and justify the imposition of 
scale invariance property (SIP) onto the estimator. 
This will be done in section 2 with various examples of scale invariant models.
In section 3 we prove a general theorem for a Bayesian estimator to be
scale invariant. This theorem states a sufficient condition on the prior laws 
which can be used for reducing the choice to admissible priors. 
For this, we consider two cases of prior laws~: 
{\em entropic prior laws} and  {\em first-order Markov models}.
In related preceding works \cite{Mohammad-Djafari90b,Mohammad-Djafari93c}, 
the SIP constraints has been studied for the case of entropic prior laws. 
In this paper we extend that work to the case of first-order Markov 
models.

\section{Linearity and scale invariance property}

In order to better understand the scale invariance property (SIP), in the next 
subsection we consider in detail the classical case of linear estimators. 
First, let us define linearity as combination of additivity:

\beq
  \forall \yb_1, \yb_2, \quad 
   \left\{\begin{array}{l}
   \yb_1 \mapsto \wh{\xb}_1 \\ 
   \yb_2 \mapsto \wh{\xb}_2
   \end{array}\right.
   \Lra   
   \yb_1 + \yb_2 \mapsto \wh{\xb}_1 + \wh{\xb}_2, 
\eeq
and the scale invariance property (SIP):

\beq
\forall \yb, \quad 
\yb \mapsto \wh{\xb}
\Lra
\forall k, \quad  k \yb \mapsto k \wh{\xb}.
\eeq
Linearity includes the SIP and so is a stronger property.
We show a particular case how the SIP is satisfied in these linear models. 

\subsection{Linearity and Gaussian assumptions}

Linear estimators under Gaussian assumptions have been (and probably still are) the
most studied Bayesian estimators because they lead to an explicit estimation formula. 
In a similar way their practical interest is  due to their easy implementation, 
such as Kalman filtering. 
In all these cases, prior laws have  the following form: 

\beq
   p_x({\xb}) \propto 
   \exp \left(-\frac{1}{2} (\xb -\mb_x)^t\Sigmab_x^{-1}(\xb -\mb_x) \right), 
\eeq
whereas the conditional additive noise is often a zero mean Gaussian process
${\cal N}(0,\Sigma_b)$.

Minimization of the posterior likelihood for all the three classical cost
functions MAP, PM and MMAP is  the  same as those of a quadratic form.
It leads to the  general form of the solution:

\beq \label{Estlineaire} 
     \widehat{\xb} =(\Ab^t\Sigmab_b^{-1} \Ab + 
                 \Sigmab_x^{-1}\Ib)^{-1}(\Ab^t\Sigmab_b^{-1}\yb +
   \Sigmab_x^{-1}\mb_x)  
\eeq
which is a linear estimator.

Some particular cases follow:

\begin{itemize}
\item Case where $\Sigmab^{-1}_x=0$ and  $\Sigmab_b=\sigma_b^2\Ib$. 
This can be interpreted as degenerated uniform prior of the solution.
The solution is the minimum variance one and is rarely suitable 
due to the high condition number of $A$.

\item Case where  $\Sigmab_b=\sigma_b^2\Ib$ and $\Sigmab_x=\sigma_x^2\Ib$.
This leads to the classical Gaussian inversion formula: 

\beq    \label{egalisation} 
\widehat{\xb} =(\Ab^t\Ab + \mu \Ib)^{-1}(\Ab^t\yb+\mu \mb_x), 
\quad\hbox{with}\quad \mu=\sigma_b^2/\sigma_x^2,
\eeq
The Signal-to-noise ratio (SNR) $\mu=\sigma_x^2/\sigma_b^2$ appears
explicitly and serves as a scale invariant parameter. 
It plays therefore the meaningful role of a hyperparameter.

\item The Gauss-Markov regularization case, which considers a smooth prior of the 
solution, is specified by setting $\Sigmab_x^{-1}= \mu \Db^t\Db+\sigma_x^{-2}\Ib$,
with $\Db$ a discrete difference matrix.
\end{itemize}

For  all these cases, estimate $\wh{\xb}$ depends on a scale.
Let us look at the dependence. 
For that matter, suppose that we change the measurement scale.
For example, if both $\xb$ and $\yb$ are optic images where each pixel 
represents the illumination (in Lumen) onto the surface of an optical device, 
we measure the number of photons coming into this device. 
(This could be of practical interest for X-ray tomography.) 
Then we convert $\yb$ into the new chosen scale and simultaneously update our parameters 
$\Sigmab_x,\Sigmab_b$~and~$\mb_x$.  
Estimation formula is then given by

\beq
     \widehat{\xb}_k =(\Ab^tk^{-2}\Sigmab_b^{-1}\Ab + 
                k^{-2} \Sigmab_x^{-1}\Ib)^{-1}(\Ab^t k^{-2}\Sigmab_b^{-1}k\yb
+ k^{-2}\Sigmab_x^{-1}k\mb_x),
\eeq
or, canceling the scale factor $k$:

\beq
     \widehat{\xb}_k =k(\Ab^t\Sigmab_b^{-1}\Ab + 
                 \Sigmab_x^{-1}\Ib)^{-1}(\Ab^t\Sigmab_b^{-1}\yb + \Sigmab_b^{-1}\mb_x).
\eeq
Thus, if we take care of  hyperparameters, the two restored images are physically the same. 

This property is rarely  stated in the Gaussian case, 
which can be explained by the use of SNR as a major tool of reasoning. 
Thus if we set the SNR, then   
$\widehat{\xb}_k$ and  $k\widehat{\xb}$ are equal. 

In many cases Gaussian assumptions are fulfilled, often leading to fast 
algorithms for calculating the resulting linear estimator.
We focus on the case where Gaussian assumptions are too strong. 
It is the case when Gauss-Markov models are used, leading to
smoother restoration than wanted. It might be explained by the 
short probability distribution tails which make discontinuity 
rare and which prevent 
appearing of wide homogeneous areas into the restored image.

\subsection{Scale invariance basics}

Although the particular case considered above may appear obvious, 
it is at the base of the scale invariance axiom.
In order to estimate or to compare physical parameters, 
we must choose a scale measurement. This can have a physical
meaningful unit or only a grey-level scale in computerized
optics. Anyway we have to keep in mind that a physical 
unit or scale is just a practical but arbitrary tool, both common and 
convenient. 
As a  consequence of this remark we state the  following axiom of scale
invariance:\\

{\it Estimation results must not depend on the arbitrary choice of the scale measurement.}
~\\

This is true when scale measurement depends on time exposure 
(astronomic observations,
Positron emission tomography, X-ray tomography, etc.).
Estimation results with two different values of time exposure must be coherent. 
SIP is also of practical interest  
when exhaustive tests are required for the validation.

Let us have a look on some regularized criteria for Bayesian estimation. 
In all the cases, the MAP criterion is used, and the estimators take the following form:

\beq  \label{mapcriteria} 
 \wh{\xb}(\yb;\psib,\lambdab) 
 = \argmins{\xb}{-\log p_b(\yb-\Ab\xb;\psib) - \log p_x(\xb;\lambdab)}.
\eeq

\medskip\noindent{\bf $L_p$--norm estimators:} 
General form of  those criteria involves an $L_p$--norm rather than a quadratic norm.
Then, the noise models and prior models take the following form:

\beq
     p_b(\yb-\Ab \xb;\psi)  \propto \expf{\psi \|\yb - \Ab \xb\|_p}
\eeq
and

\beq
    p_x(\xb;\lambda)   \propto \expf{\lambda \|\Mb \xb\|_q },
\eeq
where $\Mb$ can be a difference matrix as used by {\sc Bouman \& Sauer} and 
{\sc Besag} on the Generalized Gauss-Markov Models 
\cite{Bouman93}, and  $L_1$--Markov models  \cite{Besag89}. 
Finally, with $q=1$ and $\Mb$ an identity matrix it  
leads to a $L_1$--deconvolution algorithm in the context of seismic deconvolution 
\cite{Oldenburg86}.

According to  the scale transformation $x \mapsto kx$ and $y \mapsto
ky$, the models change in the following way:

\beq
     p_b (k \yb -A k\xb;\psi)  \propto \expf{k^p \psi \|\yb - \Ab \xb\|_p }
\eeq
and

\vspace{-6pt}
\beq
     p_x (k \xb;\lambda)  \propto \expf{k^q \lambda \|\Mb \xb\|_q }.
\eeq
If we set $(\psi_k,\lambda_k)=(k^p \psi, k^q \lambda)$, the two estimates are
scale invariant.
Moreover, if $p=q$, we can drop the scale $k$  in the
MAP criteria ({\it eq. \ref{mapcriteria}}) which becomes scale invariant.
This is done in \cite{Bouman93} \cite{Oldenburg86}, but it makes the choice
of the prior and the noise models mutually dependent.
We can also remark that $\psi^q / \lambda^p$ is scale invariant
and can be interpreted as a generalized SNR.

\medskip\noindent{\bf Maximum Entropy methods:} 
Maximum Entropy reconstruction methods have been extensively used in the last decade.
A helpful property of these methods is positivity of the restored
image. In these methods, the noise is considered zero-mean Gaussian
${\cal N}(0,\Sigma_b)$, while the Log-prior take different forms which
look like an ``Entropy measure'' of {\sc Burg} or {\sc Shannon}. 
Three different forms which have been used in practical problems are considered below. 

\begin{itemize}
 \item First, in a Fourier synthesis problem, 
{\sc Wernecke} {\it \&} {\sc D'Addario} \cite{Wernecke77} used the following form:

\beq
   p_x(\xb ;\lambda) \propto \expf{-\lambda \sum_i \log x_i }.
\eeq
Changing the scale in this context just modifies the partition function 
which is not important in the MAP criterion ({\it eq. \ref{mapcriteria}}). 
As the noise is considered Gaussian, 
these authors show that if we update the $\lambda$ parameter in a proper 
way ($\lambda_k=k^2\lambda$), then the ME reconstruction maintain linearity with respect 
to the measurement scale $k$. Thus, this ME solution is scale invariant, although nonlinear.

\item In image restoration, {\sc Burch} {\it \& al.} \cite{Burch83} , 
consider a prior law of the form

\beq
   p_x(\xb;\lambda) \propto \expf{-\lambda \sum_i x_i \log x_i }.
\eeq
Applying our scale changing yields:

\beq
   p_x(k\xb;\lambda) 
\propto \expf{-\lambda \sum_i k \, x_i \log x_i +  k \log k \, \sum_i x_i },
\eeq
which does not satisfy the scale invariance property due to the $k \log k \sum_i x_i$
term.  
It appears from their later papers that they introduced a data pre-scaling before the reconstruction.
Then, the modified version of their entropy becomes

\beq
p_x(\xb;\lambda) \propto 
 \expf{-\lambda \sum_i \frac{x_i}{s} \log\left(\frac{x_i}{s}\right) },
\eeq      
where $s$ is the pre-scaling parameter.

\item Modification of the above expression with natural parameters for
exponential family leads to the "entropic laws" used later by 
{\sc Gull \& Skilling.} \cite{Gull84} and {\sc Djafari} \cite{Mohammad-Djafari88a}:

\beq
   p_x(\xb;\lambda) 
\propto \expf{-\lambda_1 \sum_i x_i \log x_i -\lambda_2 \sum_i   x_i}.
\eeq   
The resulting estimator is scale invariant for the reasons stated above.  

\end{itemize}
   
\medskip\noindent{\bf Markovian models:} 
A new Markovian model \cite{OSullivan94} has appeared from  
$I$-divergence considerations on small translation of an image in the context 
of astronomic deconvolution. 
This model  can be rewritten as Gibbs distribution 
in the following form:

\beq
   p_x(\xb;\lambda) \propto 
  \expf{ -\lambda \sum_{(s,r) \in \cal C} (x_s-x_r) \log \left(\frac{x_s}{x_r}\right)}.
\eeq   
If we change the scale of the measurement, the scale factor $k$ vanishes in the logarithm,   
and 

\beq
   p_x(k\xb;\lambda) \propto 
 \expf{ -k\lambda \sum_{(s,r) \in {\cal C}} (x_s-x_r) \log \left(\frac{x_s}{x_r}\right)}.
\eeq 
Thus this particular Markov random field leads to a scale invariant estimator 
if we update the parameter $\lambda$ according to $\lambda \sigma_b$ constant 
(the noise is assumed Gaussian-independent).
In the same way as in the $L_p$ norm example, 
$\lambda \sigma_b$ can be considered as a generalized 
SNR.

These examples show that the family of scale invariant laws is not 
a duck-billed platypus family.
It includes many already employed priors on the context of image estimation.
We have shown in a related work that other scale invariant prior laws
exist, both in the Markovian prior family \cite{Brette94a} 
and in the uncorrelated prior \cite{Mohammad-Djafari93c} family. 

\section{Scale invariant Bayesian estimator}
Before further developing the scale invariance  constraint for the estimator, 
we want to emphasize the role of the hyperparameters $\thetab$ 
({\it i.e.}, parameters of the prior laws) and to sketch their estimation from the data 
which is very important in real-world applications. 
The estimation problem is considered globally. 
By globally we mean that, although we are interested on the estimation 
of $\xb$ we want also to take into account the estimation of the 
hyperparameters $\thetab$. 
To summarize the SIP of an estimator, we illustrate it by the following scheme:

\begin{figure} 
\begin{center} 
\hspace{-1cm}
\leavevmode
\includegraphics[width=11cm]{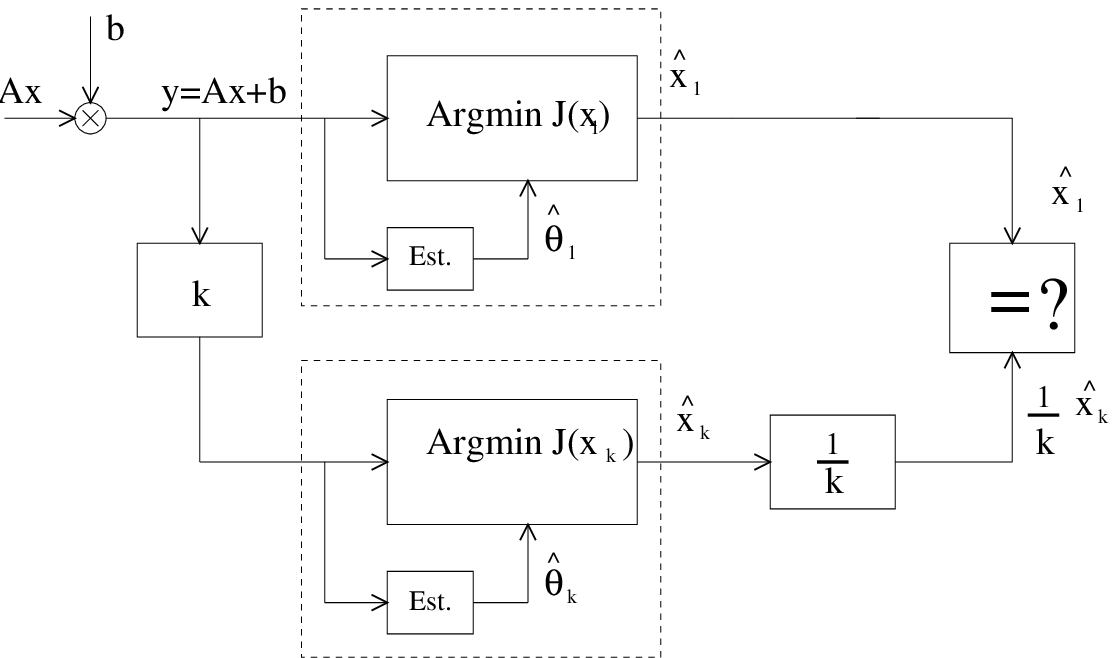}
\end{center}
\centerline{Scheme 1: Global scale invariance property for an estimator} 
\end{figure} 

For more detail, let us define a scale invariant
estimator in the following way:

\begin{Definition}      \label{ICEst}
An estimator $\wh{\xb}(\yb;\thetab)$ is said to be scale invariant 
if there exists function $\thetab_k=\fb_k(\thetab)$ such that 

\beq
\forall (\yb,\thetab,k>0), \quad \wh{\xb}(k\yb,\thetab_k)
=k \, \wh{\xb}(\yb,\thetab)
\eeq
or in short

\beq
\yb \mapsto \wh{\xb}  \Longrightarrow  
     \forall k>0, \quad k\yb \mapsto k\wh{\xb}.
\eeq
\end{Definition}

In this paper, we focus only on priors which admit density laws. We define
then the scale invariant property for those laws as follows:     

\begin{Definition} \label{ICEla}
A probability density function $p_u(\ub;\thetab)$ 
$[$resp., a conditional density $p_{u|v}(\ub|\vb;\thetab)$,$]$ 
is said to be scale invariant if there exists function $\thetab_k=\fb_k(\thetab)$ 
such that 

\beq
  \forall (\ub,\thetab,k>0),\quad p_u(k\ub;\thetab_k) =k^{-N} p_u(\ub;\thetab),
\eeq
$[$resp., \qquad $\forall (\ub,\thetab,k>0),\quad p_{u|v}(k\ub|k\vb;\thetab_k) 
=k^{-N} p_{u|v}(\ub|\vb;\thetab),]$ \\ 
where $N=\mbox{dim}(\ub)$. \\ 
If $\fb_k=Id$, \ie; if $\thetab_k=\thetab$ then $p_u(\ub;\thetab)$ is said to be strictly 
scale invariant. 
\end{Definition}

The above  property for density laws specifies that these laws are a part 
of a family of the laws which is closed relative to scale transformation.
Thus, in this class, a set of pertinent parameters exists for each chosen scale. 

We need also to set two properties for scale invariant density laws. 
Both concern the conservation of the SIP, one after marginalization,
the other after application of the Bayes rules.
 
\begin{Lemme} \label{lemme1}

If $p_{x,y}(\xb,\yb; \thetab)$ is scale invariant,  
then the marginalized $p_y(\yb;\thetab)$ is also scale invariant.
\end{Lemme}

\begin{Lemme} \label{lemme2}
 If $p_{x}(\xb; \lambdab)$ and $p_{y|x}(\yb|\xb ; \psib)$ are scale invariant,  
then the joint law  $p_{x,y}(\xb,\yb;\lambdab,\psib)$ is also scale invariant.
\end{Lemme}
Proofs are straightforward and are found in Appendix A.

Using these two definitions, we prove the following theorem which
summarizes sufficient conditions for an estimator to be scale invariant:

\begin{Theorem} \label{theorem1} 
If the cost function $C(\xb^*,\xb)$ of a Bayesian estimator satisfies the condition:

\beq
\begin{array}{l}
\forall k>0, \exists (a_k\in \ER, b_k>0) \mbox{~such that~} 
\disp{\forall (\xb^*,\xb),\quad
C(\xb_k^*,\xb_k) = a_k + b_k C(\xb^*,\xb)} ,
\end{array}
\eeq
and if the posterior law is scale invariant, {\em i.e.}, 
there exists function $\thetab_k=\fb_k(\thetab)$ such that:

\beq
\forall k>0,\forall (\xb,\yb), \quad 
p(k\xb|k\yb;\thetab_k) = k^{-\mbox{dim}(\xb)} p(\xb|\yb;\thetab), 
  \eeq
then, the resulting Bayesian estimator is scale invariant, i.e.,   

\beq
  \wh{\xb}(k\yb,\thetab_k) = k \, \wh{\xb}(\yb,\thetab).
\eeq 
\end{Theorem}

See the appendix B for the proof. It is also shown there that the cost 
functions of the three classical Bayesian estimators, 
{\em i.e.}; MAP, PM and the MMAP, satisfy the first constraint.

\medskip
\noindent{\bf Remark:} 
In this theorem, the SIP is applied to the posterior law $p(\xb|\yb;\thetab)$. 
However, we can separate the hyperparameters $\thetab$ in  two sets $\lambdab$ 
and $\psib$, where  $\lambdab$ and $\psib$ are the 
parameters of the prior laws $p_x(\xb;\lambdab)$ and $p_b(\yb-\Ab \xb;\psib)$.  
In what follows, we want to make the choice of $p_x$ and $p_b$ independent.  
From the lemma 1 and 2, if  $p_x$ and $p_b$ satisfy the SIP then the posterior 
$p(\xb|\yb;\thetab)$ satisfies the SIP. As a consequence $\thetab_k$ must be separated 
according to $\thetab_k=[\lambdab_k,\psib_k]=[\gb_k(\lambdab), \hb_k(\psib)]$. 

\section{Hyperparameters estimation} 
In the above theorem, we assumed that the hyperparameters $\thetab$ 
are given. Thus, given the data $\yb$ and the hyperparameters $\thetab$, 
we can calculate $\wh{\xb}$. Now, if the scale factor $k$ of the data 
has been changed, we have first to update the hyperparameters \cite{Marroquin93}
according to 
$\thetab_k=\fb_k(\thetab)$, and then we can use the SIP: 

\beq
\wh{\xb}(k\yb,\thetab_k) = k \, \wh{\xb}(\yb,\thetab).
\eeq
Now, let us  see what happens if we have to estimate both $\xb$ and 
$\thetab$, either by Maximum or Generalized Maximum Likelihood. 

\begin{itemize}
\item Maximum likelihood (ML) method estimates first $\thetab$ by 

\beq
  \wh{\thetab}=\argmaxs{\thetab}{L(\thetab)} ,
\eeq
where 

\vspace{-6pt}
\beq \label{likelihood}
L(\thetab)= p(\yb ;\thetab)
\eeq
and then $\wh{\thetab}$ is used to estimate $\xb$.
At a scale $k$,

\beq
   \wh{\thetab}_k=\argmaxs{\thetab_k}{L_k(\thetab_k)}.
\eeq
Application of lemma \ref{lemme1} implies that

\beq
   L_k(\thetab_k)=k^{\mbox{dim}(\yb)} L(\thetab),
\eeq
thus, the Maximum Likelihood estimator satisfies the condition

\beq
\wh{\thetab}_k=\fb_k(\wh{\thetab}).
\eeq
 
The likelihood function ({\it eq.} \ref{likelihood}) has rarely an explicit form, and  
a common algorithm for its locally maximization is the EM algorithm which is an 
iterative algorithm described briefly as follows: 

\beq
\left\{
 \begin{array}{ccc} 
 Q(\thetab;\wh{\thetab}^{(i)}) 
 & = & \espx{\xb|\yb;\wh{\thetab}^{(i)}}{\ln p(\yb|\xb;\thetab)}\\
 \wh{\thetab}^{(i+1)} 
 & = & \disp{\argmax{\thetab}{Q(\thetab;\wh{\thetab}^{(i)})}}.
 \end{array}
\right.
\eeq
At a scale $k$,  

\beqn
 Q_k(\thetab_k;\wh{\thetab}_k^{(i)}) 
  &=& \espx{k\xb|k \yb;\wh{\thetab}_k^{(i)}}{\ln p(k\yb|k\xb;\thetab_k)} \nonumber\\
  &=& -M \ln k +\espx{k \xb|k \yb; \wh{\thetab}_k^{(i)}}{\ln p(\yb|\xb;\thetab)} \nonumber\\
  &=& -M \ln k +k^{-{\mbox{dim}(\yb)}}  Q(\thetab;\wh{\thetab}^{(i)}).
\eeqn
Thus, if we initialize this iterative algorithm with the value 
$\wh{\thetab}_k^{(0)} = f_k(\wh{\thetab}^{(0)})$, 
then we have 

\beq
   \wh{\thetab}_k^{(1\cdots l)} = \fb_k\left(\wh{\thetab}^{(1\cdots l)}\right).
\eeq
Then the scale invariance coherence of hyperparameters is ensured 
during the optimization steps.
 
\item In Generalized Maximum Likelihood (GML) method, one estimates 
both $\thetab$ and $\xb$ by 

\beq
\left(\wh{\thetab},\wh{\xb}\right)=\argmaxs{(\thetab,\xb)}{p(\xb,\yb;\thetab)}.
\eeq
Applying the same demonstration as above to the joint laws rather than to the marginalized
one leads to 

\beq
\left(\wh{\thetab}_k,\wh{\xb}_k\right) = \left(\fb_k(\wh{\thetab}),k\wh{\xb}\right).
\eeq
However, this holds if and only if the GML has a maximum. 
This may not be always the case and this is a major drawback in GML. 
Also, in GML method, direct resolution is rarely possible and sub-optimal techniques lead to the
classical two-step estimation scheme:

\beqn
\wh{\xb}^{(i)}        & = & \argmaxs{\xb}{p(\xb,\yb;\wh{\thetab}^{(i)})}, \\
\wh{\thetab}^{(i+1)} & = & \argmaxs{\thetab}{p(\wh{\xb}^{(i)},\yb;\thetab)}.
\eeqn

We see that, in each iteration, the $\thetab $ estimation step may be considered as the ML
estimation of $\thetab$ if $\xb^{(i)}$ is supposed to be a realization of the 
prior law.  Thus the coherence of estimated hyperparameters at different scales is fulfilled 
during the both optimization steps, and 

\beq
  \left( \wh{\thetab}_k^{(1\cdots l)},\wh{\xb}_k^{(1\cdots l)} \right)
  =\left( \fb_k(\wh{\thetab}^{(1\cdots l)}), k \wh{\xb}^{(1\cdots l)} \right).
\eeq
\end{itemize}

Thus, if  we consider the whole estimation problem (with a ML or GML approach), 
the SIP of the estimator is assured in both cases. It is also ensured
during the iterative optimization schemes of ML or GML. 

%\section{Entropic scale invariant laws}

\section{Markovian invariant distributions}

Markovian distributions as priors in image processing 
allow to introduce local characteristics and inter-pixels correlations. 
They are widely used but there exist many different Markovian models and 
very few model selection guidelines exist.  
In this section we apply the above scale invariance considerations 
to the prior model selection in the case of first order homogeneous MRFs.

Let  $X \in \Omega$ be a homogeneous Markov random field  defined on the subset 
$[1 \ldots N] \times[1 \ldots M]$ of  $\EZ^2$.
The Markov characteristic property is:

\beq
    p_X(x_i|x_{ {\cal S}-i}) = p_X(x_i|x_{\delta i}),
\eeq
where  $\delta i$ is the neighbourhood of site $i$, and  $\cal S$ is the
set of pixels.
Hammersley-Clifford theorem for the first order neighbourhood reads:

\beq
  p_X(\xb;\lambda) \propto \mbox{exp} \left( -\lambda \sum_{\{r,s\} \in {\cal C}} 
               \phi(x_s,x_r)     \right),
\eeq  
where  $\cal C$ is the clique set, and $ \phi(x,y) $
the clique potential. In most works 
\cite{Bouman93,Geman84,Geman92,Besag86} a simplified model is introduced under the form 
$\phi(x,y) =\phi(x- y)$. Here we keep a general point of view. 
Application of  the scale invariance condition 
to the Markovian prior laws $p_X(\xb,\lambda)$ leads to the two following theorems:

\begin{Theorem} 
A familly of Markovian distribution is scale invariant 
if and only if there exist two functions $f(k,\lambda)$ 
and $\beta(k)$ such that clique potential $\phi(x_s,x_r) $
satisfies:

\beq \label{eq5}
 f(k,\lambda)\ \phi(k x_s,k x_r) =  \lambda \phi(x_s,x_r) +\beta(k).
\eeq
\end{Theorem}

\begin{Theorem}
A necessary and sufficient condition for a Markov random fields to be scale invariant is 
that exists a triplet $(a,b,c)$ such as the clique potential $\phi(x_s,x_r)$  
verifies the linear partial differential equation (PDE) ~:
\[
a \phi(x_s,x_r) + b \left( x_s \dpdx{\phi(x_s,x_r)}{x_s} + 
     x_r \dpdx{\phi(x_s,x_r)}{x_r} \right) = c.
\]
\end{Theorem}

Finally, enforcing symmetry of the clique potentials 
$\phi(x_s,x_r) =\phi(x_r,x_s)$ the following theorem provides the set of  
scale invariant clique potentials:

\begin{Theorem}
$p_X(\xb,\lambda)$ is  scale invariant if and only if  
$\phi(x_s,x_r)$ is chosen from one of the following vector spaces:     

\beq \label{ICE1} 
{\cal V}_0 =
\biggl\{ \phi(x_s,x_r) \mid \exists  \psi(.) \hbox{~even~and~} p \in \ER, 
     \phi(x_s,x_r) = \psi \left( \log  \left| 
         \frac{x_s}{x_r} \right| \right) - p \log |x_s x_r| 
              \biggr\}   
\eeq
\beq \label{ICE2}
{\cal V}_1(p) =
\biggl\{ \phi(x_s,x_r) \mid \exists \psi(.)   \hbox{~even~} , \phi(x_s,x_r) =
 \psi \left( \log \left| \frac{x_s}{x_r} \right| \right)  |x_s x_r|^p \biggr\}
\eeq
Moreover, ${\cal V}_0$ is the subspace of strictly scale invariant clique potentials.
\end{Theorem} 
For the proof of these theorems see \cite{Brette94b}.  

Among the most common models in use for image processing purposes, only few clique
potentials fall into the above set. Let us give two examples:

First, the GGMRFs proposed by {\sc Bouman \& Sauer} \cite{Bouman93} were built 
by a similar approach of scale invariance but under the 
restricted assumption that $\phi(x_s,x_r) =\phi(x_s-x_r)$. The yielded expression  
$\phi(x_s,x_r) =|x_s-x_r|^p$ can be factored according to 
$\phi(x_s,x_r)=|x_s\, x_r|^{p/2} |\mbox{2sh} \, (\log(x_s/x_r)/2)|^{p}$ which shows 
that it falls in ${\cal V}_1(p)$.

The  second example of potential does not reduce to the single variable function \\  
$\phi(x_s-x_r)$:
$\phi(x_s,x_r)=(x_s-x_r)\log{\left( x_s/x_r \right)}$. It has recently been  
introduced  from  I-divergence penalty considerations
in the field of image estimation problem (optic deconvolution)
by O'Sullivan \cite{OSullivan94}. 
  Factoring  $|x_s
x_r|^{\frac{1}{2}}$ leads to: 

\beq 
    \phi(x_s,x_r)=|x_s x_r|^{\frac{1}{2}} \psi\left( \log(x_s/x_r) \right),
 \eeq
where  $\psi(X) =2X \mbox{sh} (X/2)$ is even.  It shows that
$\phi(x_s,x_r)$  is in ${\cal V}_1(1/2)$ and is scale invariant.
 As $\phi(x_s,x_r)$ is defined
only on $\ER^2_{*+}$ it applies to 
positive quantities. This feature is very useful  in image processing
where prior positivity applies to many physical quantities.

\section{Conclusions}

In this paper we have outlined and justified a weaker property than linearity 
that is desired for the Bayesian estimators to have. 
We have shown that this scale invariance
property (SIP) helps to avoid an arbitrary choice for the scale of the measurement.   
Some models already employed  in Bayesian estimation, 
including Markov prior Models \cite{Bouman93,OSullivan94}, Entropic prior
\cite{Gull89,Mohammad-Djafari93c} and Generalized Gaussian models
\cite{Oldenburg86},  
have demonstrated the existence and usefulness of scale invariant models.  
%In some cases, the SIP plays the meaningful role of extended SNR as a regularization parameter.  
Then we have given general conditions for a Bayesian estimator to be
scale invariant. 
This property holds for most Bayesian estimators such as MAP, PM, MMAP 
under the condition that the prior laws are also scale invariant. 
Thus, imposition of the SIP can assist in the model
selection. We have also shown that classical hyperparameters estimation methods 
satisfy the SIP property for estimated laws. 

Finally we discussed how to choose the prior laws to obtain scale 
invariant Bayesian estimators. For this, we considered two cases: 
{\em entropic prior laws} and  {\em first-order Markov models}.
In related preceding works \cite{Mohammad-Djafari90b,Mohammad-Djafari93c,Mohammad-Djafari93d}, 
the SIP constraints have been studied for the case of entropic prior laws. 
In this paper we extended that work to the case of first-order Markov 
models and showed that many common Markov models used in image processing 
are special cases.  

\appendix
\section{SIP property inheritance }

\begin{itemize}
\item {\bf Proof of the Lemma \ref{lemme1}:}
  
Let $ p_{x,y}(\xb,\yb; \thetab) $  have the scale invariance property, then
if there exists
$\thetab_k=\fb_k(\thetab)$ such that
\[
  p_{x,y}(k\xb,k\yb; \thetab_k) =  k^{-(M+N)} p_{x,y}(\xb,\yb; \thetab),
\]
where $N = \mbox{dim}(\xb)$ and $M = \mbox{dim}(\yb)$, 
then, marginalizing with respect to $\xb$, we obtain
\[
p_y(k\yb;\thetab_k)
 = k^{-(M+N)} \intg p_{x,y}(\xb,\yb; \thetab) k^{-N} d{\xb}  
 = k^{-M}  p_y(\yb; \thetab),
\]
which completes the proof.

\item {\bf  Proof of the Lemma \ref{lemme2}:}  

The definition of SIP for density laws and direct
application of the Bayes rule lead to

\[
p_{x,y}(k\xb,k\yb; \thetab_k) 
 =  k^{-N} p_{x}(\xb; \lambdab) \, k^{-M} p_{y|x}(\yb|\xb; \psib)
 =  k^{-(M+N)} p_{x,y}(\xb,\yb; \thetab), 
\]
which concludes the proof.
\end{itemize}

\section{SIP conditions for Bayesian estimator}

\begin{itemize}
\item {\bf Proof of the Theorem \ref{theorem1}:} 

Since a Bayesian estimator is defined by
\[ 
\wh{\xb}
=\argmins{\xb}{\int C(\xb^*,\xb) \,  p(\xb^*|\yb;\thetab) \,d{\xb^*}},
\] 
then

\vspace{-18pt}
\begin{eqnarray*}
\wh{\xb}_k 
&=&\argmins{\xb_k}{\int C(\xb_k^*, \xb_k) \, p(\xb_k^*|k\yb;{\thetab}_k) d(\xb_k^*)} \\
&=& k \,\argmins{\xb}{\int C(k\xb^*,k\xb) \, p(k\xb^*|k\yb;\thetab_k) \, k^N \d{\xb^*}} \\
&=& k \,\argmins{\xb}{\int [a_k + b_k C(\xb^*,\xb)] k^{-N} \, 
     p(\xb^*|\yb;\thetab) \, k^N \d{\xb^*}} 
= k \, \wh{\xb},
\end{eqnarray*} 
which proves the Theorem \ref{theorem1}.

\item {\bf Conditions for cost functions}: 

The three classical Bayesian estimators, MAP, PM and MMAP, 
satisfy the condition of the cost function:  
\begin{itemize} 
\item {\it  Posterior mean} (PM):\quad
\( 
C(\xb_k^*,\xb_k) 
= (\xb_k^*-\xb_k)^t  \Qb \, (\xb_k^*-\xb_k)
= k^2 \, C(\xb^*,\xb).
\)

\item  {\it Maximum a posteriori}~(MAP):\quad 
\(
C(\xb_k^*,\xb_k) = 1-\delta(\xb_k^*-\xb_k)= C(\xb^*,\xb).
\)

\item  {\it Marginal Maximum a Posteriori}~(MMAP):
\[ 
C(\xb_k^*,\xb_k) 
= \sum_i \left( 1-\delta([\xb_k^*]_i-[\xb_k]_i) \right)
= C(\xb^*,\xb).
\]
\end{itemize}
\end{itemize}

% fichier : revuedef.tex
% Auteur: AMD, GLB
% Mise a jour: 06 septembre 1992, 10:00
%
% Modifie et completee par	L.Lemitre le 05 10 94
% 				J.-F. Gio le 15-02-95
%				H. C. le 27-06-95
%				J. I. le 02-08-95
%
% Ce fichier contient des alias sur les noms de revues
% pour faciliter la saisie de fiches biblio.
%
\def\AA{Astrononmy and Astrophysics}
\def\AAP{Advances in Applied Probability}			%Ajout FC
\def\ABE{Annals of Biomedical Engineering}
\def\AT{Annales des T\'el\'ecommunications}
\def\AMS{Annals of Mathematical Statistics}
\def\AISM{Annals of Institute of Statistical Mathematics}
\def\AO{Applied Optics}
\def\AP{The Annals of Probability}
\def\AST{The Annals of Statistics}
\def\BMK{Biometrika}
\def\CPAM{Communications on Pure and Applied Mathematics}
\def\EMK{Econometrica}
\def\CRAS{Compte-rendus de l'acad\'emie des sciences}
\def\CVGIP{Computer Vision and Graphics and Image Processing}
\def\GJRAS{Geophysical Journal of the Royal Astrononomical Society}
\def\GSC{Geoscience}
\def\GPH{Geophysics}
\def\GRETSI#1{Actes du #1$^{\mbox{e}}$ Colloque GRETSI} 
%\de%\GRETSI89{Actes du 12 Colloque GRETSI} 
%\de%\GRETSI91{Actes du 13 Colloque GRETSI} 
%\de%\GRETSI93{Actes du 14 Colloque GRETSI} 
%\de%\GRETSI95{Actes du 15 Colloque GRETSI} 
\def\CGIP{Computer Graphics and Image Processing}
\def\ICASSP{Proceedings of IEEE ICASSP}
\def\ICEMBS{Proceedings of IEEE EMBS}
\def\ICIP{Proceedings of the International Conference on Image Processing}
\def\ieeP{Proceedings of the IEE}
\def\ieeeAC{IEEE Transactions on Automatic and Control}
\def\ieeeAES{IEEE Transactions on Aerospace and Electronic Systems}
\def\ieeeAP{IEEE Transactions on Antennas and Propagation}
\def\ieeeASSP{IEEE Transactions on Acoustics Speech and Signal Processing}
\def\ieeeBME{IEEE Transactions on Biomedical Engineering}
\def\ieeeCS{IEEE Transactions on Circuits and Systems}
\def\ieeeCT{IEEE Transactions on Circuit Theory}
\def\ieeeC{IEEE Transactions on Communications}
\def\ieeeGE{IEEE Transactions on Geoscience and Remote Sensing}
\def\ieeeGEE{IEEE Transactions on Geosciences Electronics}
\def\ieeeIP{IEEE Transactions on Image Processing}
\def\ieeeIT{IEEE Transactions on Information Theory}
\def\ieeeMI{IEEE Transactions on Medical Imaging}
\def\ieeeMTT{IEEE Transactions on Microwave Theory and Technology}
\def\ieeeM{IEEE Transactions on Magnetics}
\def\ieeeNS{IEEE Transactions on Nuclear Sciences}
\def\ieeePAMI{IEEE Transactions on Pattern Analysis and Machine Intelligence}
\def\ieeeP{Proceedings of the IEEE}
\def\ieeeRS{IEEE Transactions on Radio Science}
\def\ieeeSMC{IEEE Transactions on Systems, Man and Cybernetics}
\def\ieeeSP{IEEE Transactions on Signal Processing}
\def\ieeeSSC{IEEE Transactions on Systems Science and Cybernetics}
\def\ieeeSU{IEEE Transactions on Sonics and Ultrasonics}
\def\ieeeUFFC{IEEE Transactions on Ultrasonics Ferroelectrics and Frequency Control}
\def\IJC{International Journal of Control}
\def\IJCV{International Journal of Computer Vision}
\def\IJIST{International Journal of Imaging Systems and Technology}
\def\IP{Inverse Problems}
\def\ISR{International Statistical Review}
\def\IUSS{Proceedings of International Ultrasonics Symposium}
\def\JAPH{Journal of Applied Physics}
\def\JAP{Journal of Applied Probability}
\def\JAS{Journal of Applied Statistics}
\def\JASA{Journal of Acoustical Society America}
\def\JASAS{Journal of American Statistical Association}
\def\JBME{Journal of Biomedical Engineering}			% Ajout JFG
\def\JCAM{Journal of Computational and Applied Mathematics}	% Ajout HC
\def\JCAT{Journal of Computer Assisted Tomography}
\def\JEWA{Journal of Electromagnetic Waves and Applications}	% Ajout HC
\def\JMO{Journal of Modern Optics}
\def\JNDE{Journal of Nondestructive Evaluation}		        % Ajout HC
\def\JMP{Journal of Mathematical Physics}
\def\JOSA{Journal of Optical Society America}
\def\JP{Journal de Physique}
\def\JRSSA{Journal of the Royal Statistical Society A}
\def\JRSSB{Journal of the Royal Statistical Society B}
\def\JRSSC{Journal of the Royal Statistical Society C}
\def\JSPI{Journal of Statistical Planning and Inference}        %Ajout MF
\def\JTSA{Journal of Time Series Analysis}                      %Ajout MF
\def\JVCIR{Journal of Visual Communication and Image Representation} 
	\def\MMAS{???} % Trouve dans gpi base 
\def\MNAS{Mathematical Methods in Applied Science}
\def\MNRAS{Monthly Notes of Royal Astronomical Society}
\def\MP{Mathematical Programming}
	\def\NSIP{NSIP}  % Trouve dans gpi base 
\def\OC{Optics Communication}
\def\PRA{Physical Review A}
\def\PRB{Physical Review B}
\def\PRC{Physical Review C}
\def\PRD{Physical Review D}				% Ajout stef
\def\PRL{Physical Review Letters}			% Ajout HC	
\def\RGSP{Review of Geophysics and Space Physics}	% Ajout HC			% Ajout HC	
\def\RS{Radio Science}					% Ajout HC	
\def\SP{Signal Processing}
\def\siamAM{SIAM Journal of Applied Mathematics}
\def\siamCO{SIAM Journal of Control and Optimization}
\def\siamJO{SIAM Journal of Optimization}		% Ajout HC pour JFB
\def\siamMA{SIAM Journal of Mathematical Analysis}
\def\siamNA{SIAM Journal of Numerical Analysis}
\def\siamO{SIAM Journal of Optimization}
\def\siamR{SIAM Review}
\def\SSR{Stochastics and Stochastics Reports}           % Ajout MF
\def\TPA{Theory of Probability and its Applications}	% Ajout FC
\def\TMK{Technometrics}
\def\TS{Traitement du Signal}
\def\UMB{Ultrasound in Medecine and Biology}
\def\US{Ultrasonics}
\def\USI{Ultrasonic Imaging}

\bibliographystyle{ieeetr}
\bibliography{gpibase}

\end{document}